\pdfoutput=1

\documentclass[10pt]{article}
\usepackage{graphicx}

\def\Title#1{\begin{center} {\Large #1 } \end{center}}
\def\Author#1{\begin{center}{ \sc #1} \end{center}}
\def\Address#1{\begin{center}{ \it #1} \end{center}}

\newcommand\pubblock{\rightline{\begin{tabular}{l} Proceedings of the Second Annual LHCP\\ \pubnumber\\
         \pubdate  \end{tabular}}}

\newenvironment{Abstract}{\begin{quotation} \begin{center} 
             \large ABSTRACT \end{center}\bigskip 
      \begin{center}\begin{large}}{\end{large}\end{center} \end{quotation}}

\newenvironment{Presented}{\begin{quotation} \begin{center} 
             PRESENTED AT\end{center}\bigskip 
      \begin{center}\begin{large}}{\end{large}\end{center} \end{quotation}}

\def\beq{\begin{equation}}
\def\eeq#1{\label{#1}\end{equation}}
\def\eeqn{\end{equation}}

\def\beqa{\begin{eqnarray}}
\def\eeqa#1{\label{#1}\end{eqnarray}}
\def\eeqan{\end{eqnarray}}

\let\bar=\overbar

\def\Dslash{\not{\hbox{\kern-4pt $D$}}}
\def\dslash{\not{\hbox{\kern-2pt $\del$}}}

\def\msb{{\bar{\ssstyle M \kern -1pt S}}}

\usepackage{color}

\newcommand\pubnumber{ }

\newcommand\pubdate{\today}

\def\affiliation{
On behalf of the ATLAS Collaboration, \\
Department of Physics \\
Oxford University, Oxford OX1 3RH, UK }

\begin{document}
\large
\begin{titlepage}
\pubblock

\vfill
\Title{ ATLAS measurements of multi-boson production }
\vfill

\Author{ Christopher Hays  }
\Address{\affiliation}
\vfill
\begin{Abstract}

Measurements of electroweak gauge-boson pair-production in $\sqrt{s}=7$ and 8 TeV $pp$ 
collisions at the LHC probe self-couplings and interference effects to an accuracy of 
${\cal{O}}(10\%)$ or better.  ATLAS measurements of $ZZ$ and $WZ$ production at both 
center of mass energies, and of $WW$, $Z\gamma$ and $W\gamma$ production at 
$\sqrt{s}=7$~TeV, are presented.  Total, fiducial, and differential cross sections 
are given, along with limits on anomalous triple-gauge couplings.

\end{Abstract}
\vfill

\begin{Presented}
The Second Annual Conference\\
 on Large Hadron Collider Physics \\
Columbia University, New York, U.S.A \\ 
June 2-7, 2014
\end{Presented}
\vfill
\end{titlepage}
\def\thefootnote{\fnsymbol{footnote}}
\setcounter{footnote}{0}
%

\normalsize 

\section{Introduction}

The study of vector-boson pair-production at the LHC provides important 
information on gauge-boson self-couplings and on QCD corrections, which 
are particularly relevant for Higgs boson measurements in these final states.  
Measurements of diboson ($VV$) production at ATLAS~\cite{ATLAS} generally 
include a cross section measured in the selected fiducial region, an 
extrapolation to a total production cross section, a cross section measured 
differentially in relevant variables, and limits on anomalous couplings.  

The fiducial cross section is defined as 
\begin{equation}
\sigma_{fid} = \frac{N_{data} - N_{bd}}{{\cal{L}}C_{VV}},
\end{equation}

\noindent
where $N_{data}$ is the number of observed events, $N_{bd}$ is the number of 
events expected from background, ${\cal{L}}$ is the integrated luminosity, and 
$C_{VV}$ is the expected ratio of all selected events to the generated events 
in a relevant fiducial region.  To extrapolate to the total cross section, the 
fiducial cross section is divided by $A_{VV}$, the ratio of events in the 
fiducial region to all generated events, and ${\cal{B}}$, the relevant branching 
ratios.  

Diboson measurements probe triple-gauge couplings, which contain a charged current in 
the SM.  The general Lagrangian terms that conserve $C$ and $P$ for these couplings 
are~\cite{atgc}
\begin{equation}
{\cal{L}}_{WWV} = ig_1^V(W^{\dagger}_{\mu\nu} W^{\mu} V^{\nu} - W^{\dagger}_{\mu} V_{\nu} 
W^{\mu\nu}) + i\kappa_V W^{\dagger}_{\mu} W_{\nu} V^{\mu\nu} + \frac{i\lambda_V}{m_W^2}
W^{\dagger}_{\lambda\mu} W_{\nu}^{\mu} V^{\nu\lambda}, 
\end{equation}

\noindent
where $g_1^{V,SM} = \kappa^{SM}_V = 1$ and $\lambda^{SM}_V = 0$ in the standard model 
(SM).  The parameter $g_1^{\gamma}$ is fixed to 1 by electromagnetic gauge 
invariance.  Anomalous coupling limits are set in three scenarios: ``LEP'', where 
$\Delta \kappa_{\gamma} = (\cos^2\theta_W/\sin^2\theta_W)(\Delta g_1^Z - \Delta\kappa_Z)$ 
and $\lambda_Z=\lambda_{\gamma}$, leaving three parameters; ``HISZ'', where 
$\Delta g_1^Z = \Delta\kappa_Z/(\cos^2\theta_W - \sin^2\theta_W)$, 
$\Delta\kappa_{\gamma} = 2\Delta\kappa_Z\cos^2\theta_W/(\cos^2\theta_W - \sin^2\theta_W)$, 
and $\lambda_Z=\lambda_{\gamma}$, leaving two parameters; and ``equal couplings'', 
where $\Delta\kappa_{\gamma}=\Delta\kappa_Z$, $g_1^Z=1$, and $\lambda_Z=\lambda_{\gamma}$, 
leaving two parameters.  Generally the anomalous couplings lead to a violation of 
unitarity, so a suppression factor of, e.g., 
$\lambda(\hat{s}) = \lambda/(1+\hat{s}/\Lambda^2)^2$ is applied. 

In order of increasing rate, ATLAS has measured $ZZ$ production in four-lepton data 
in 7 and 8 TeV collisions; $WZ$ production in three-lepton data at both energies; 
$WW$ production in dilepton and mono-lepton data at 7 TeV; $Z\gamma$ production in 
dilepton-plus-photon and monophoton data at 7 TeV; $W\gamma$ production in 
lepton-plus-photon data at 7 TeV; and $\gamma\gamma$ production at 7 TeV.

\section{$ZZ$ cross sections}

\begin{figure}[htpb]
\centering
\includegraphics[height=2.5in]{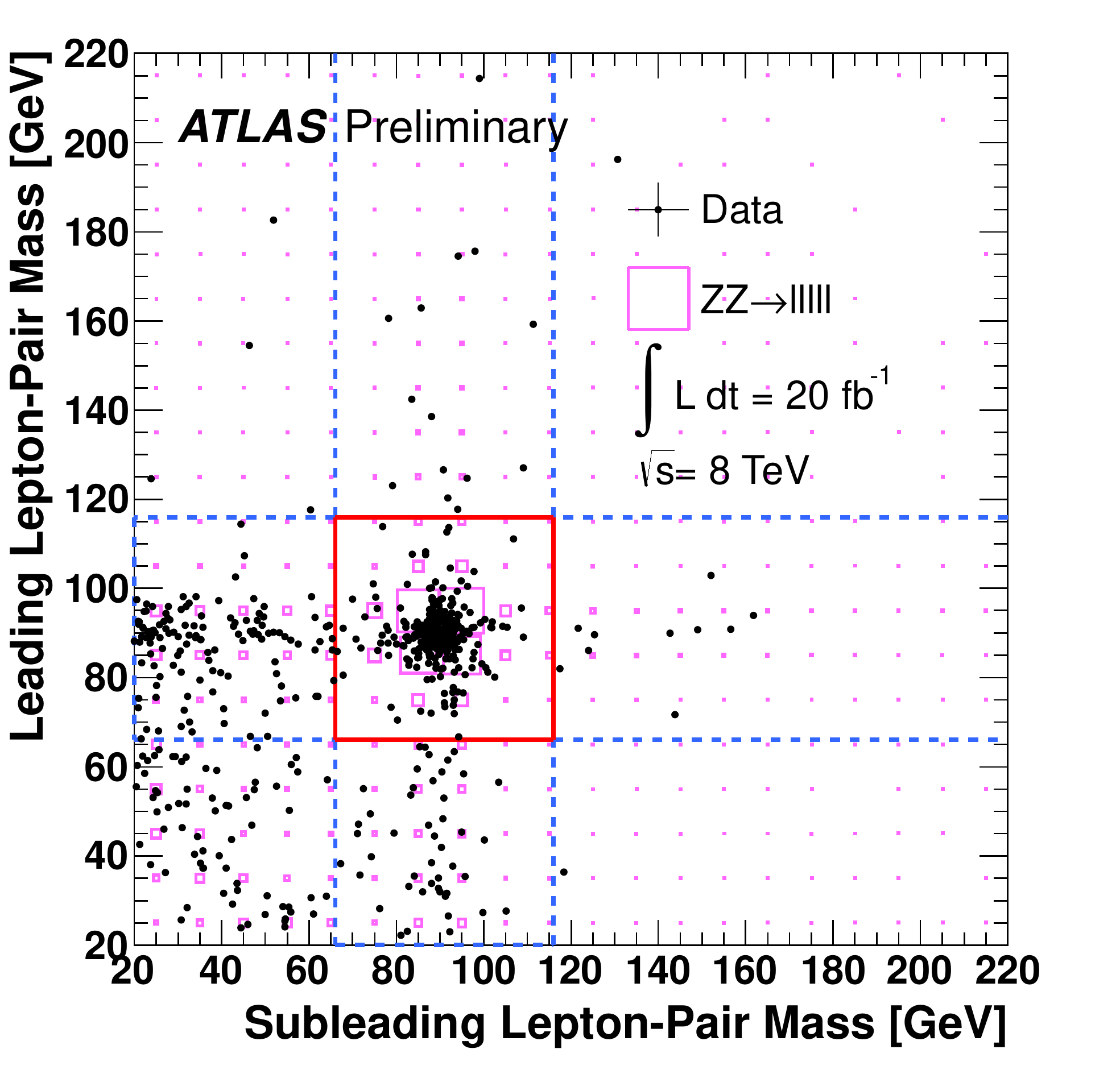}
\includegraphics[height=2.5in]{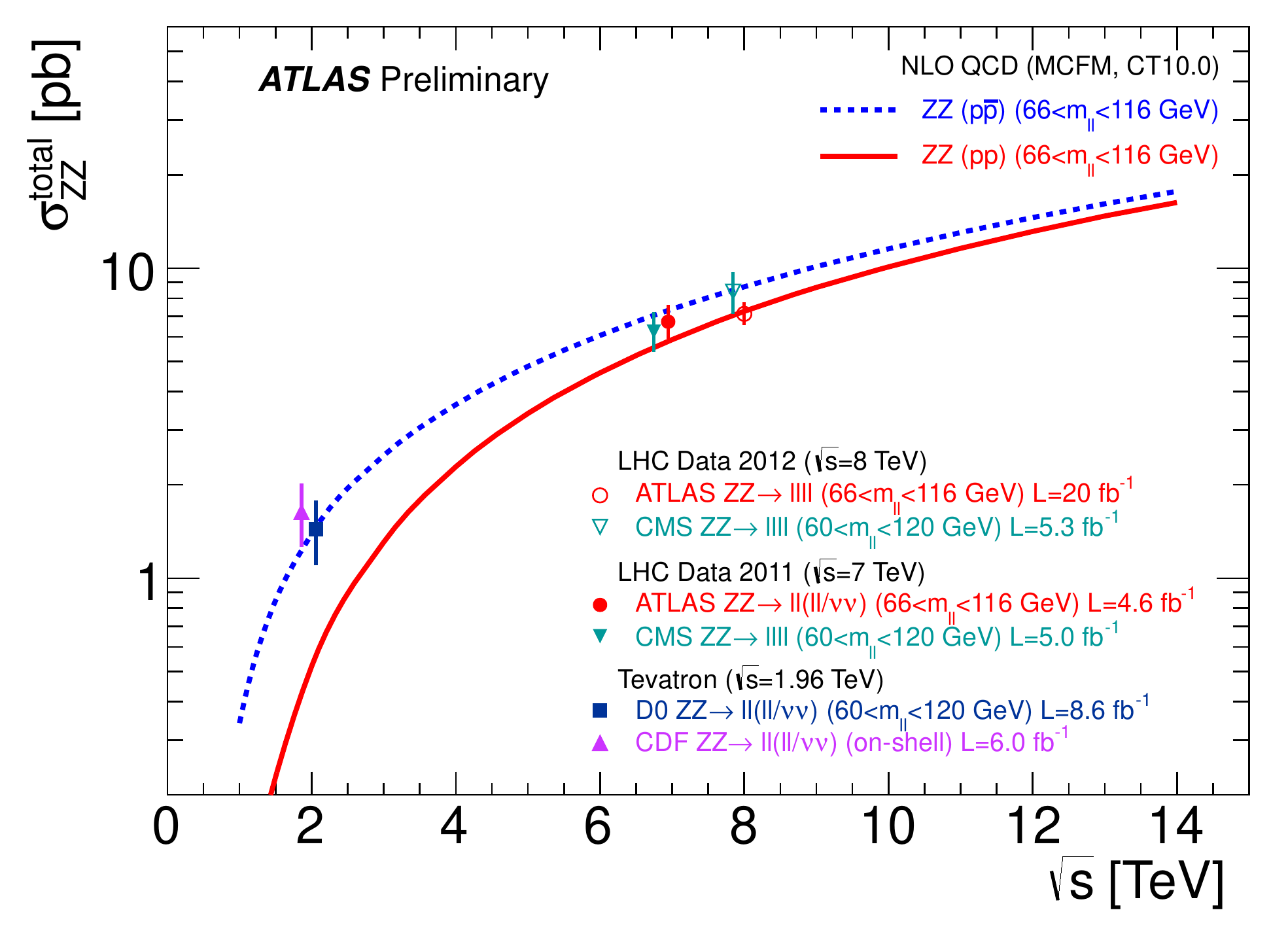}
\caption{Left: Invariant masses of the leading and subleading lepton pairs in 
events with two pairs of opposite-charge electrons or muons.  The leading lepton 
pair has higher $p_T$~\cite{ZZ8TeV}.  Right: The measured $ZZ$ cross section as a 
function of center of mass energy for $p\bar{p}$ and $pp$ collisions~\cite{ZZ8TeV}. }
\label{fig:zzplots}
\end{figure}

For the $ZZ$ cross section measurements at $\sqrt{s}=8$ TeV~\cite{ZZ8TeV}, candidate 
events are selected with two pairs of opposite-charge electrons or muons with invariant 
mass consistent with $m_Z$ ($66 < m_{ll} < 116$ GeV).  Figure~\ref{fig:zzplots} shows 
the distribution of the invariant mass of the pair with the highest dilepton $p_T$ (the 
``leading'' pair) versus the invariant mass of the ``subleading'' pair with lower 
dilepton $p_T$.  

This selection provides $>90\%$ purity, with $292.5 \pm 10.6$ expected $ZZ$ events 
compared to a background of $20.4 \pm 5.8$ events in $20.3 \pm 0.6$~fb$^{-1}$ of 
integrated luminosity.  With 305 observed events, the measured total cross section 
$\sigma_{tot}(ZZ) = 7.1^{+0.5}_{-0.4}~{\rm (stat.)} \pm 0.8~{\rm (sys.)} \pm 
0.2~{\rm (lum.)}$~pb is consistent with the SM prediction ($7.2^{+0.3}_{-0.2}$~pb).  
The measured fiducial cross section is 
$\sigma_{fid}(ZZ\rightarrow llll) = 20.7^{+1.3}_{-1.2}~{\rm (stat.)} 
\pm 0.8~{\rm (sys.)} \pm 0.6~{\rm (lum.)}$~fb.

Including prior measurements, there is good agreement with SM predictions for $ZZ$ 
production in hadron collisions at center of mass energies ranging from 2 to 8 TeV 
(Fig.~\ref{fig:zzplots}).  Measurements in $pp$ collisions have been somewhat higher 
than the expectation, but not significantly so.
 
\section{$WZ$ cross sections}
\begin{figure}[htpb]
\centering
\includegraphics[height=2.25in]{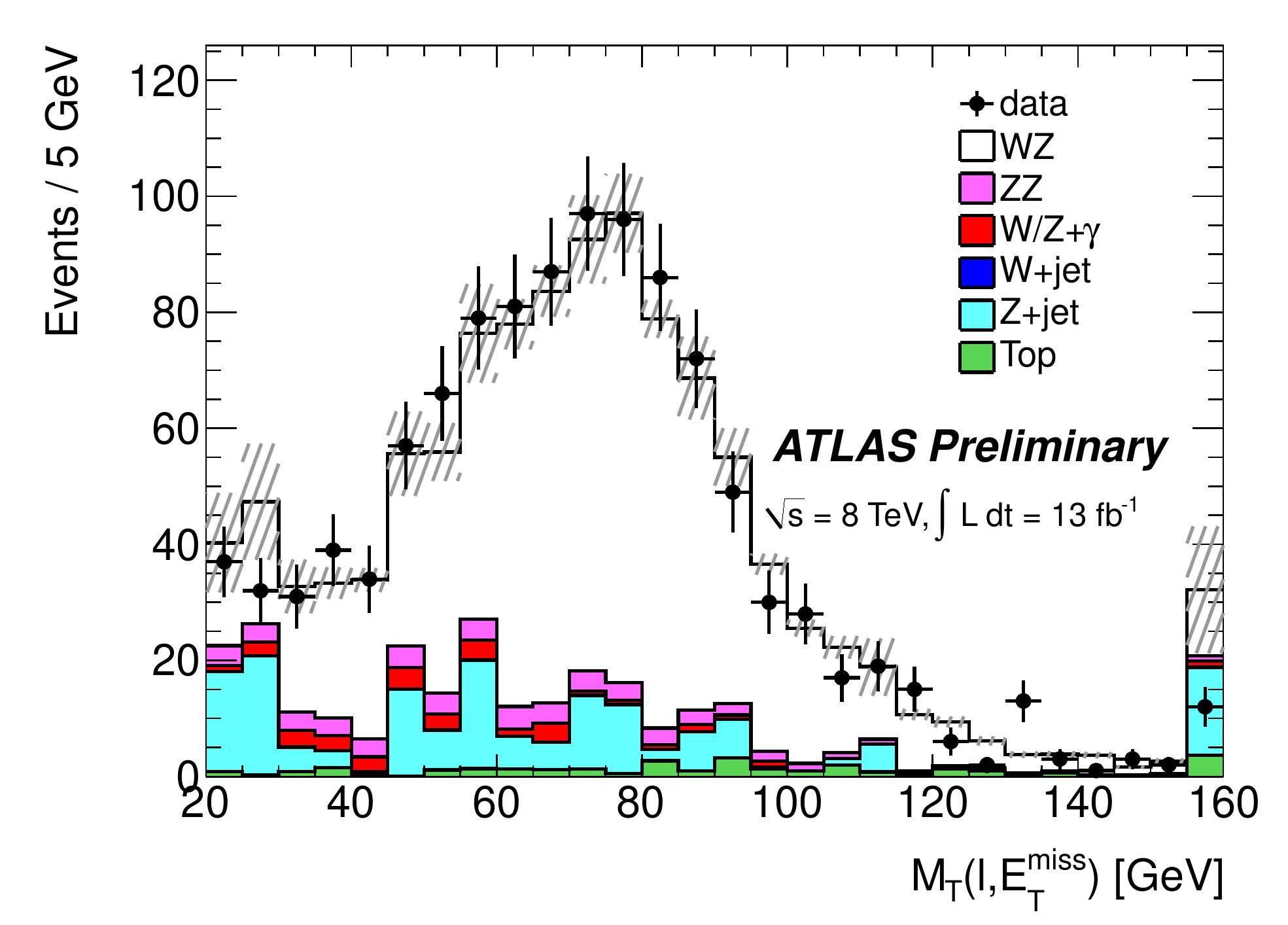}
\includegraphics[height=2.25in]{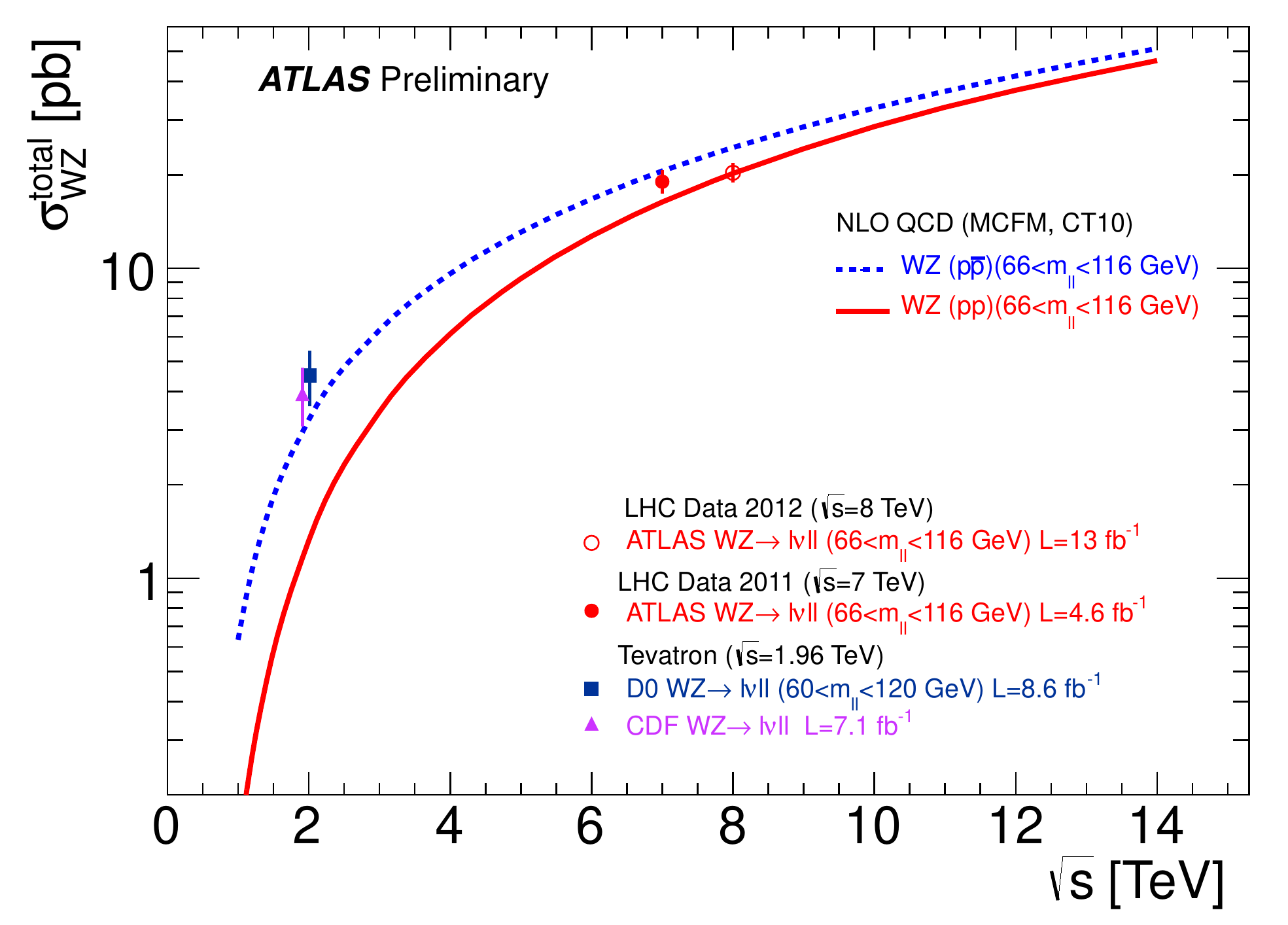}
\caption{Left: Transverse mass $m_T$ calculated using the lepton not associated with 
the $Z$-boson decay, and the reconstructed missing transverse momentum~\cite{WZ}.  
Right:  The measured $WZ$ cross section as a function of center of mass energy for 
$p\bar{p}$ and $pp$ collisions~\cite{WZ}. }
\label{fig:wzplots}
\end{figure}

Using data corresponding to 13~fb$^{-1}$ of integrated luminosity at 
$\sqrt{s} = 8$~TeV~\cite{WZ}, candidate $WZ$ events are selected by requiring: 
an opposite-charge electron or muon pair with invariant mass near $m_Z$ 
($81 < m_{ll} < 101$ GeV); an additional electron or muon; and significant 
missing transverse momentum ($E_T^{miss} > 25$~GeV) in a direction that is not 
collinear with the additional lepton ($m_T > 20$ GeV, where 
$m_T = \sqrt{2(E_T^{\ell}E_T^{\nu} - \vec{p}_T^{~\ell}\vec{p}_T^{~\nu})}$).  
With this selection, the sample is 75\% pure in $WZ$ events, with $819\pm34$ 
expected signal events and $277 \pm 26$ background events.  The 1094 observed 
events are consistent with the expectation for $WZ$ production, as demonstrated with 
the $m_T$ distribution in Fig.~\ref{fig:wzplots}.  Combining all lepton channels, the 
measured fiducial cross section has 7\% precision, $\sigma_{fid}(WZ\rightarrow l\null) = 
99.2^{+3.8}_{-3.0}~{\rm (stat.)} ^{+5.1}_{-5.4}~{\rm (sys.)}^{+3.1}_{-3.0}~{\rm (lum.)}$~fb, 
and is equal to the SM expectation ($99.2 \pm 3.6$~fb).  The total measured cross 
section is $\sigma_{tot}(WZ) = 20.3^{+0.8}_{-0.7}~{\rm (stat.)}^{+1.2}_{-1.1}~{\rm (sys.)}
^{+0.7}_{-0.6}~{\rm (lum.)}$~pb, again equal to the SM expectation ($20.3\pm0.8$~pb).  
As with the $ZZ$ cross section measurements, the $WZ$ measurements are consistent 
with SM predictions for $\sqrt{s}$ ranging from 2-8 TeV for $p\bar{p}$ and $pp$ 
collisions (Fig.~\ref{fig:wzplots}).

\section{$WW$ measurements}

In 4.6 fb$^{-1}$ of $\sqrt{s}=7$~TeV data, ATLAS has measured $WW$ cross sections both 
inclusively and differentially as a function of the $p_T$ of the highest momentum lepton 
in the event~\cite{WW}.  Events are selected with two opposite-charge leptons (electron 
or muon); same-flavor leptons must have invariant mass inconsistent with $m_Z$ 
($|m_{ll} - m_Z| > 15$ GeV).  The top-quark background is suppressed by requiring 
events to have zero jets, and Drell-Yan processes are reduced by requiring significant 
$E_T^{miss,rel}$, defined as $E_T^{miss}$ multiplied by the sine of the smallest 
$\Delta\phi(\vec{p}_T^{~\ell},\vec{p}_T^{~\nu})$ when $\Delta\phi < \pi/2$.  

The $WW$ selection results in $824 \pm 69$ expected signal $WW$ events and $369 \pm 61$ 
background events, for a purity of nearly 70\%.  Because of the reduced Drell-Yan 
background in events with different-flavor leptons, these events contain nearly 
two-thirds of the signal acceptance.  The measured fiducial cross section in this 
final state is $\sigma_{fid}(WW\rightarrow e\nu\mu\nu) = 262.3 \pm 12.3~{\rm (stat.)} 
\pm 20.7~{\rm (sys.)} \pm 10.2~{\rm (lum.)}$~fb.  Combining with same-flavor final 
states, the total cross section is 
$51.9 \pm 2.0~{\rm (stat.)} \pm 3.9~{\rm (sys.)} \pm 2.0~{\rm (lum.)}$~pb, which is a 
little more than $1\sigma$ higher than the prediction of $44.7^{+2.1}_{-1.9}$~pb.  

\begin{figure}[tpb]
\centering
\includegraphics[height=2.4in]{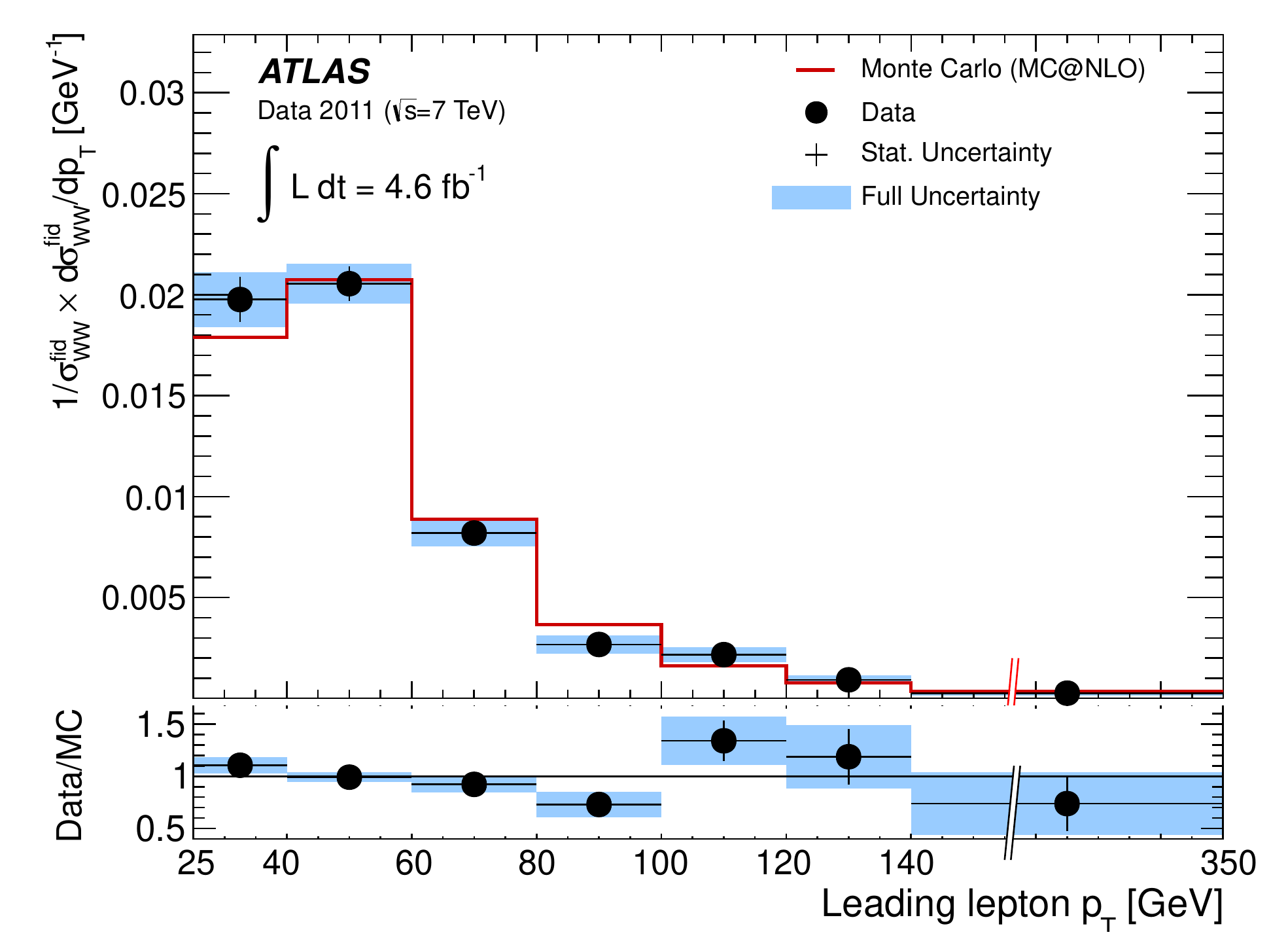}
\includegraphics[height=2.4in]{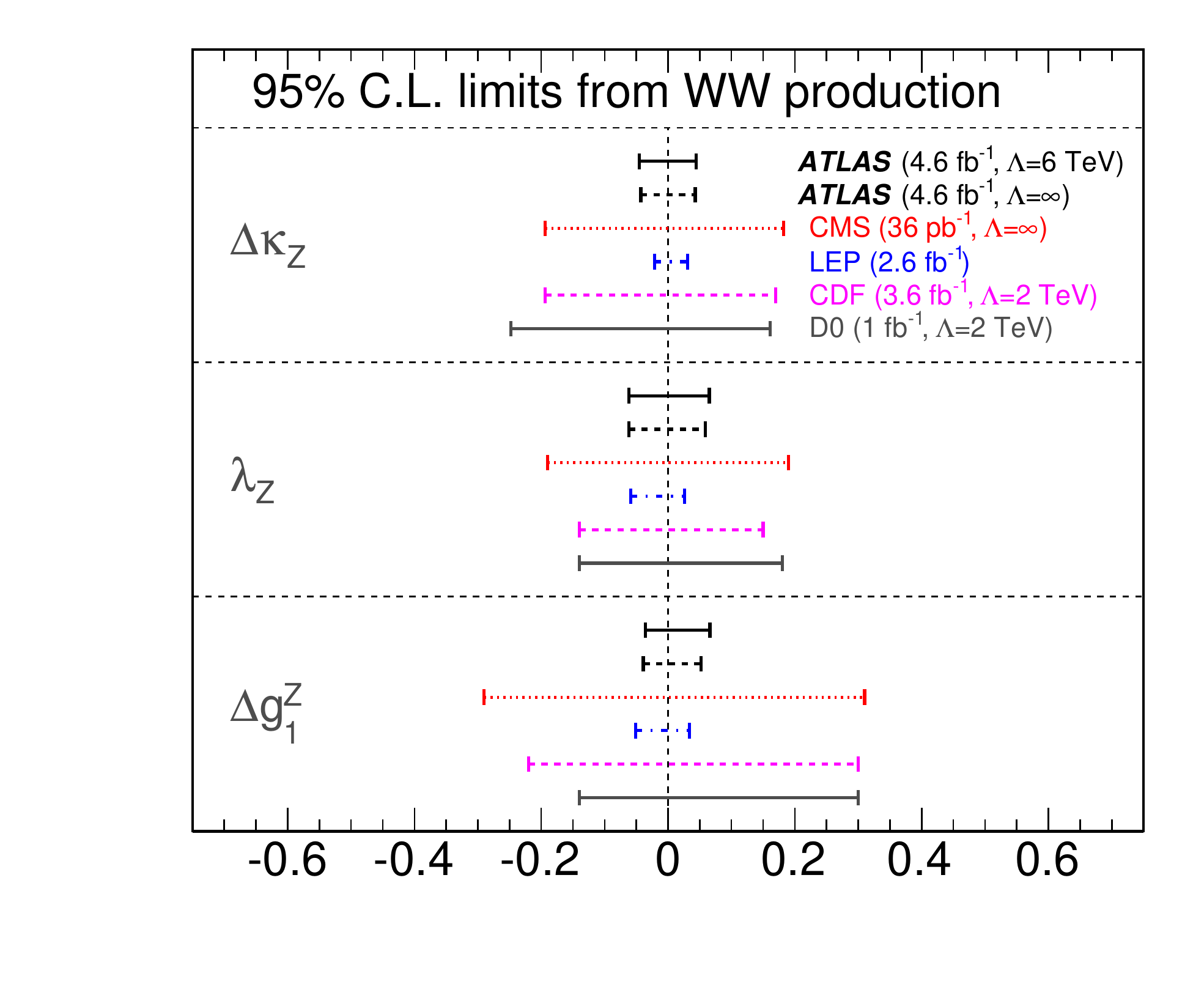}
\caption{Left: Differential cross section for $WW$ production as a function of leading 
lepton $p_T$, normalized by the total cross section~\cite{WW}.  Right: Limits on anomalous 
gauge couplings from the fit to the reconstructed distribution of leading lepton $p_T$ 
(ATLAS), along with limits from other experiments (CMS, LEP, CDF, D0)~\cite{WW}. } 
\label{fig:wwplots}
\end{figure}

The $p_T$ of the lepton with the highest $p_T$ (the ``leading'' lepton) probes the $Q^2$ 
of the event, and is sensitive to anomalous gauge couplings.  The cross section is 
measured differentially as a function of leading lepton $p_T$, with results shown in 
Fig.~\ref{fig:wwplots}.  The reconstructed leading lepton $p_T$ is fit for the presence 
of anomalous gauge couplings and limits are set in the absence of evidence for these 
couplings.  Figure~\ref{fig:wwplots} shows the resulting limits for $\Lambda=6$~TeV, 
which preserves unitarity for coupling values that are not excluded, and for 
$\Lambda=\infty$.  Limits from prior measurements are shown for comparison. 

\section{$Z\gamma$ measurements}

Measurements of $Z\gamma$ production have been performed by ATLAS for both 
$Z\rightarrow \ell\ell$ and $Z\rightarrow\nu\nu$ decays, using 4.6~fb of integrated 
luminosity from $\sqrt{s}=7$~TeV collisions~\cite{WZgamma}.  Events from the decay 
to charged leptons are selected by requiring two same-flavor leptons (electrons or 
muons) with invariant mass larger than 40~GeV and a photon with $p_T > 15$~GeV.  
The selection yields 3990 expected signal events (with $\approx 5\%$ uncertainty) 
and 677 background events (with $\approx 30\%$ uncertainty), for a purity of 85\%. 
For decays to neutrinos, the large background from photon plus jet production is 
suppressed by requiring a photon with $p_T>100$ GeV and $E_T^{miss} > 90$ GeV.  
The direction of $E_T^{miss}$ must be opposite the photon 
[$\Delta\phi(E_T^{miss},\gamma)>2.6$] and not in the direction of a jet 
[$\Delta\phi(E_T^{miss},jet)>0.4$].  Background from $W$-boson production is 
suppressed by removing events with an identified electron or muon.  This selection 
yields 420 signal events (with $\approx 15\%$ uncertainty) and 670 background 
events (with $\approx 10\%$ uncertainty), for a purity of just under 40\%.  The 
purity is increased to 55\% by requiring the events to have no reconstructed jets.  
The $Z\gamma$ measurements are performed both inclusively in jets and exclusively 
using events with no jets.

\begin{figure}[tb]
\centering
\includegraphics[height=2.3in]{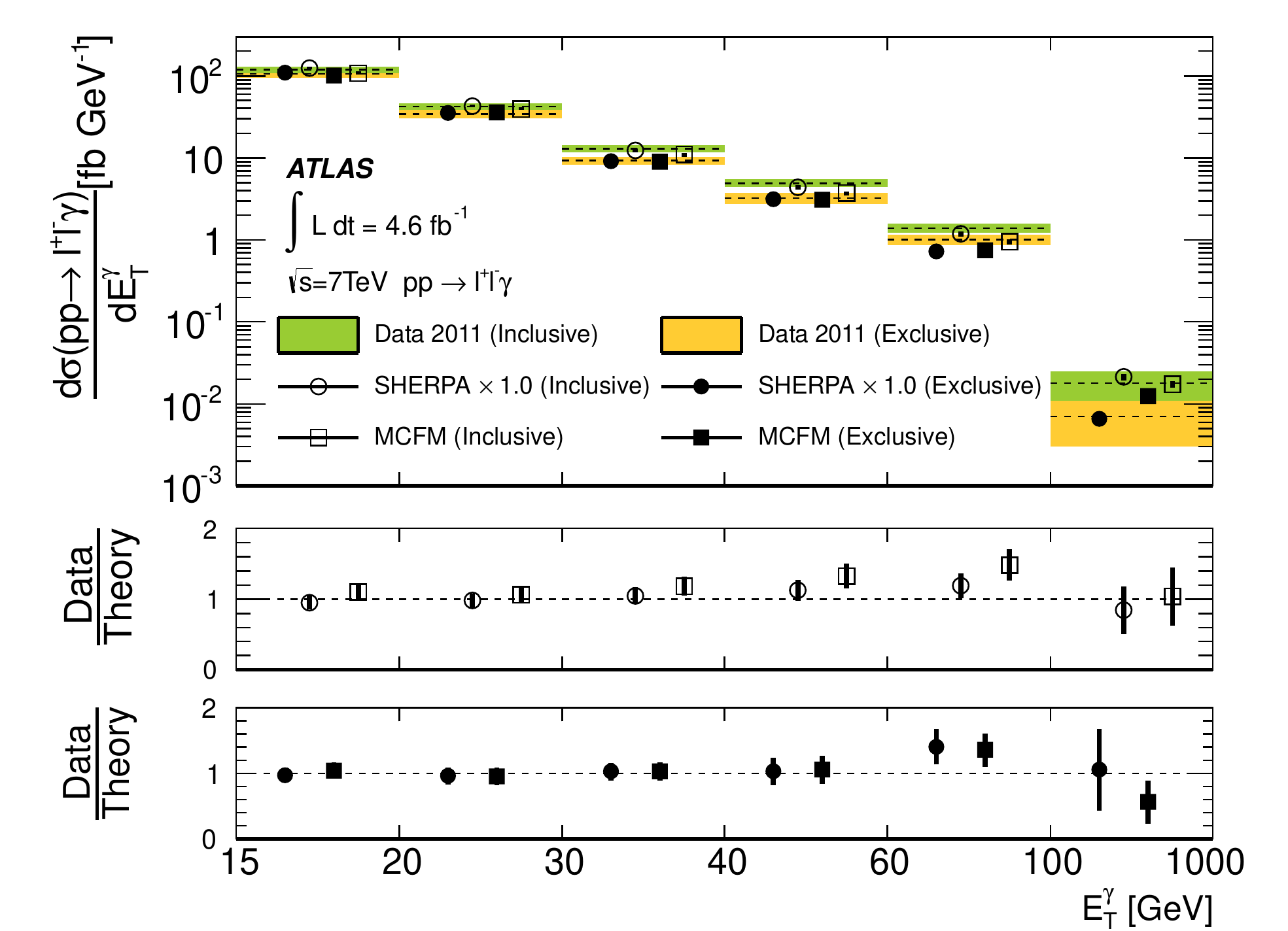}
\includegraphics[height=2.3in]{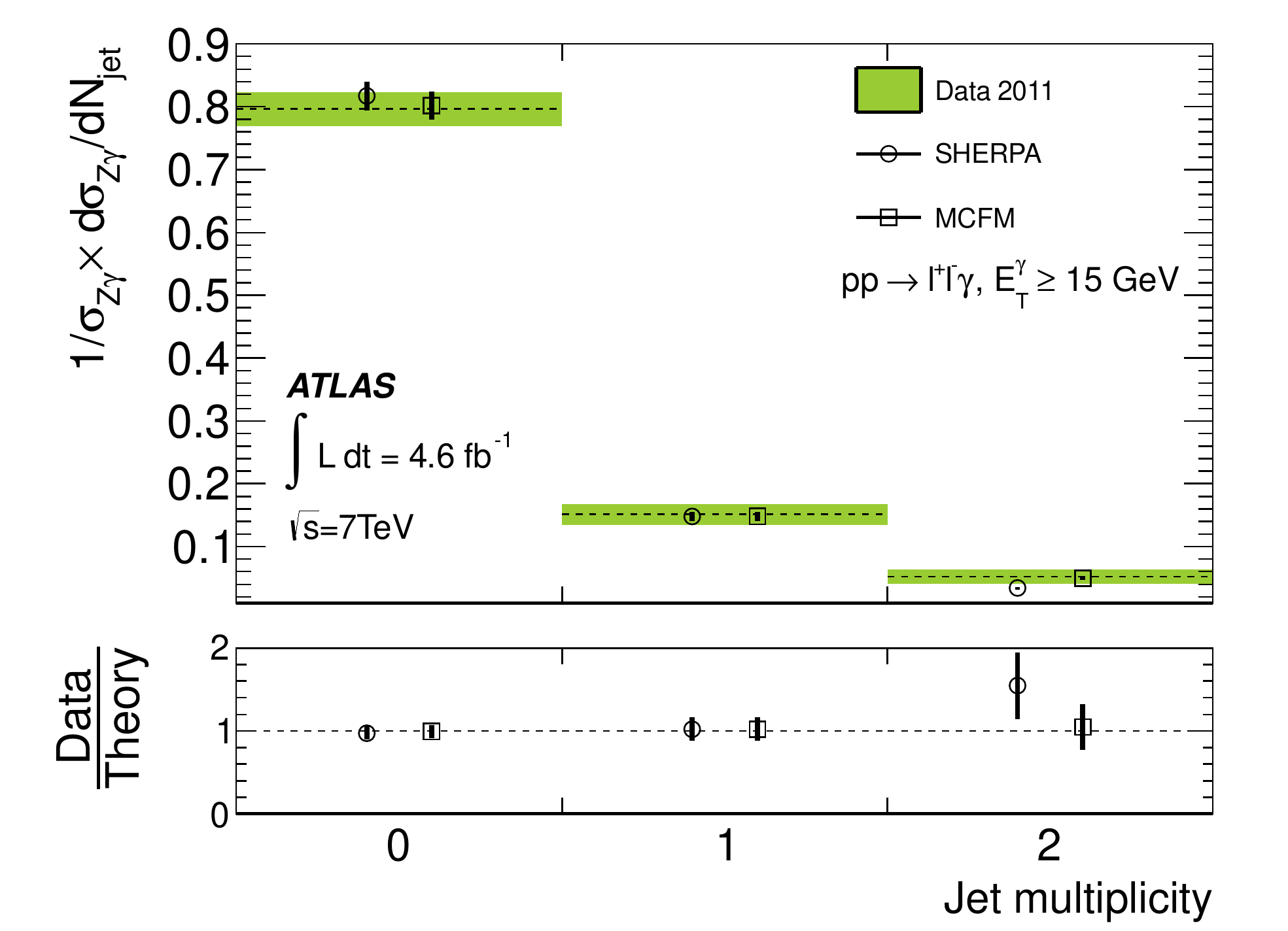}
\caption{The differential cross sections for $Z\gamma\rightarrow\ell\ell\gamma$ 
production as functions of photon $E_T$ (left) and jet multiplicity 
(right)~\cite{WZgamma}. }
\label{fig:zgplots}
\end{figure}

The measured $Z\gamma$ fiducial cross sections are consistent with the MCFM 
predictions, as shown in Table~\ref{tab:Zgamma}.  In the $\ell\ell\gamma$ final 
state, the cross section is measured differentially in jet multipicity, photon 
$E_T$, and invariant mass of the $\ell\ell\gamma$ system.  Figure~\ref{fig:zgplots} 
shows good agreement between the measurement and the predictions of MCFM or Sherpa, 
though MCFM tends to underestimate the rate of events with photons at high $E_T$.  
These can arise from higher-order production of quarks that radiate a high-momentum 
photon.  Sherpa is generated with tree-level diagrams of up to three additional 
quarks or gluons (partons), whereas MCFM has at most one additional parton in this 
distribution.  

Events with a reconstructed photon with $E_T>100$~GeV are used to set limits 
on anomalous $ZZ\gamma$ or $Z\gamma\gamma$ vertices, with limits significantly 
improved over those of CDF and D0 at the Tevatron. 

\begin{table}[tb]
\begin{center}
\begin{tabular}{ccc}  
 Final state & Measured $\sigma_{fid}$~(pb) & $\sigma_{fid}^{SM}$~(pb) \\
\hline
 $\ell\ell\gamma$ ($\geq 0$ jets) & $1.30 \pm 0.03~{\rm (stat.)} \pm 0.11~{\rm (sys.)} \pm 0.05~{\rm (lum.)}$
&  $1.18 \pm 0.05$      \\
 $\ell\ell\gamma$ (0 jets) &  $1.05 \pm 0.02~{\rm (stat.)} \pm 0.10~{\rm (sys.)} \pm 0.04~{\rm (lum.)}$
&  $1.06 \pm 0.05$       \\
\hline
 $\nu\nu\gamma$  ($\geq 0$ jets) &  $0.133 \pm 0.013~{\rm (stat.)} \pm 0.020~{\rm (sys.)} \pm 0.005~{\rm (lum.)}$
&  $0.156 \pm 0.012$     \\ 
 $\nu\nu\gamma$  (0 jets) &  $0.133 \pm 0.013~{\rm (stat.)} \pm 0.020~{\rm (sys.)} \pm 0.005~{\rm (lum.)}$     
&  $0.115 \pm 0.009$     \\ 
\hline
\end{tabular}
\caption{ The measured and predicted $Z\gamma$ cross sections in the $\ell\ell\gamma$ 
and $\nu\nu\gamma$ final states, for events inclusive in jets and events with no 
jets~\cite{WZgamma}. }
\label{tab:Zgamma}
\end{center}
\end{table}

\section{$W\gamma$ measurements}

\begin{figure}[tb]
\centering
\includegraphics[height=2.5in]{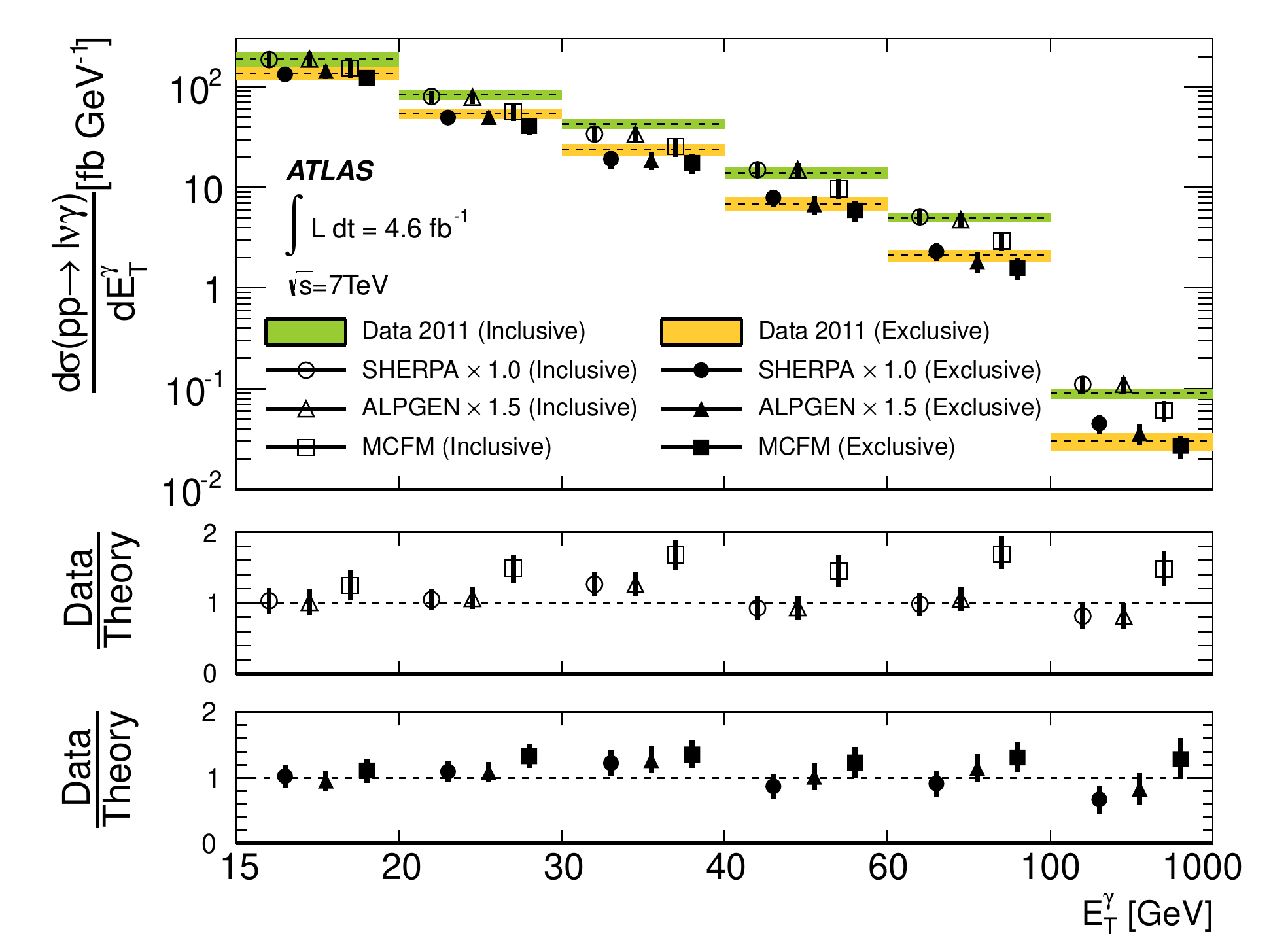}
\includegraphics[height=2.5in]{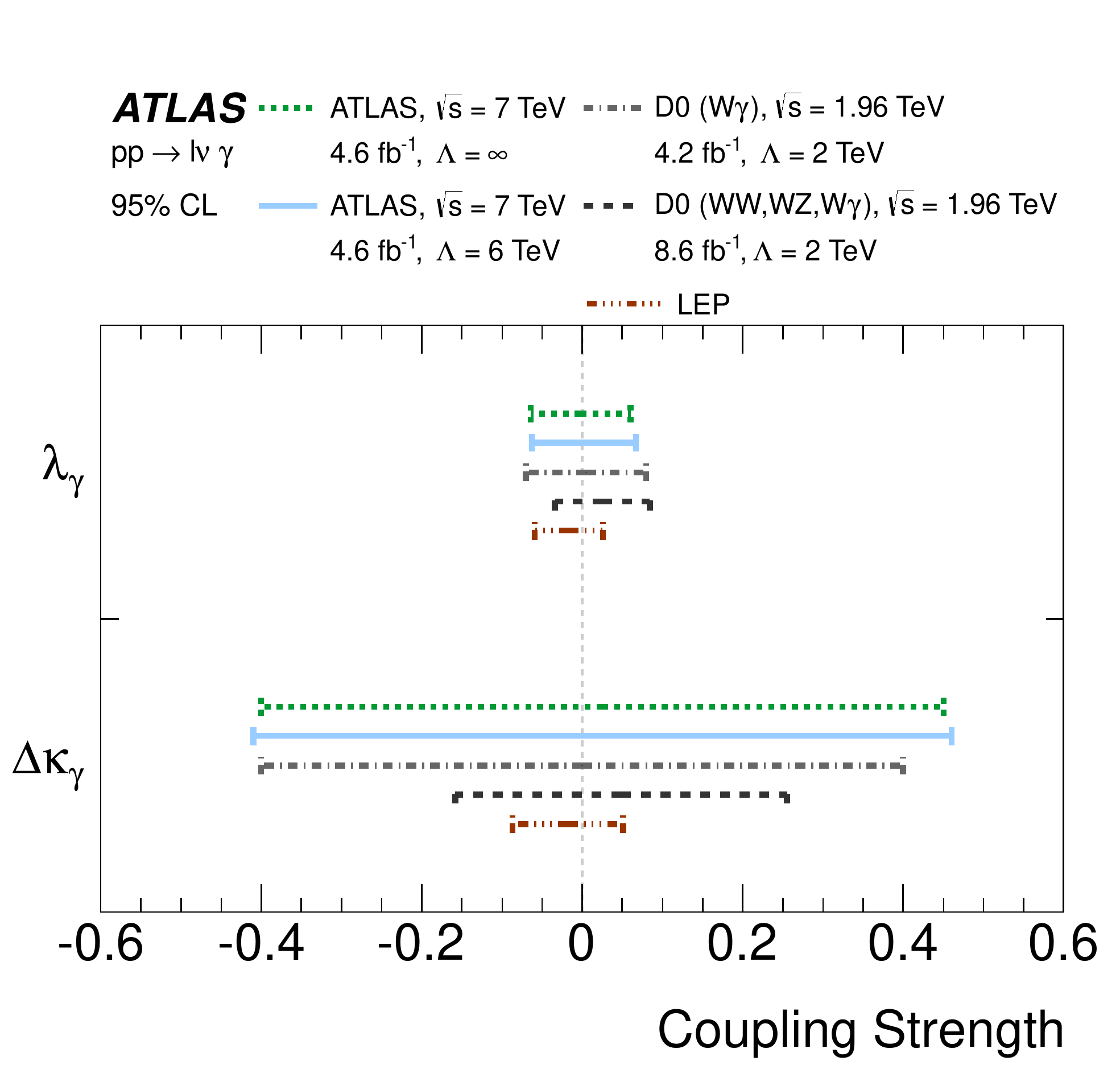}
\caption{Left: The differential $W\gamma\rightarrow\ell\nu\gamma$ cross section as a function of 
photon $E_T$~\cite{WZgamma}.  Right:  Limits on anomalous couplings compared to those from D0 and 
LEP~\cite{WZgamma}. }
\label{fig:wgplots}
\end{figure}

The highest rate of diboson production is $W\gamma$, producing more than 10000 signal 
events after event selection.  The selection requires an electron or muon, a photon, and 
large $E_T^{miss}$, giving a purity of $\approx 60\%$.  As with the $Z\gamma$ measurements, 
fiducial cross sections are measured inclusively in jets and in events with no reconstructed 
jets (Table~\ref{tab:Wgamma}).  The measured inclusive cross section is more than $2\sigma$ 
higher than the MCFM prediction, due to the upper limit of one additional partons produced 
by MCFM.  The agreement with MCFM is improved in events with no reconstructed jets, though 
the cross section measurement is still higher than the prediction.  The differential cross 
section with respect to photon $E_T$ (Fig.~\ref{fig:wgplots}) shows a discrepancy with MCFM 
that increases with increasing photon $E_T$, as with the $Z\gamma$ measurement.  

Anomalous couplings are probed using events with photon $E_T>100$~GeV.  The resulting limits 
on $\lambda_{\gamma}$ and $\Delta\kappa_{\gamma}$ are shown in Fig.~\ref{fig:wgplots} for 
$\Lambda=6$~TeV and $\Lambda=\infty$, assuming all other couplings have their SM values.  

\begin{table}[tb]
\begin{center}
\begin{tabular}{ccc}  
 Final state & Measured $\sigma_{fid}$~(pb) & $\sigma_{fid}^{SM}$~(pb) \\
\hline
 $\ell\nu\gamma$ ($\geq 0$ jets) & $2.77 \pm 0.03~{\rm (stat.)} \pm 0.33~{\rm (sys.)} \pm 0.14~{\rm (lum.)}$
&  $1.96 \pm 0.17$      \\
 $\ell\nu\gamma$ (0 jets) &  $1.76 \pm 0.03~{\rm (stat.)} \pm 0.21~{\rm (sys.)} \pm 0.08~{\rm (lum.)}$
&  $1.39 \pm 0.13$       \\
\hline
\end{tabular}
\caption{ The measured and predicted $W\gamma$ cross sections in the $\ell\nu\gamma$ 
final state, for events inclusive in jets and events with no jets~\cite{WZgamma}. }
\label{tab:Wgamma}
\end{center}
\end{table}

\section{Summary}

ATLAS has measured diboson cross sections in $\sqrt{s}=7$ and 8 TeV, including a number of unfolded 
differential cross sections.  The cross sections are sensitive to higher order perturbative QCD 
predictions, and demonstrate the importance of including multiple partons in the predictions. 
Anomalous coupling limits have been set using the results from $\sqrt{s}=7$ TeV data.

\end{document}